\documentclass{jfm}
\usepackage{graphicx,subfigure,natbib}
\usepackage{amssymb,amsmath}

\def\Sp{{\rm Sp}}

\newcommand\bs{\hat{\mathbf{s}}}
\newcommand\bx{\hat{\mathbf{x}}}
\newcommand\bz{\hat{\mathbf{z}}}
\newcommand\rmd{\mathrm{d}}
\newcommand\bu{{\mathbf{u}}}
\newcommand\bF{{\mathbf{F}}}
\newcommand\bT{{\mathbf{T}}}
\newcommand\bM{{\mathbf{M}}}
\newcommand\bq{{\mathbf{q}}}

\title[Shape of optimal active  flagella]
{Shape of  optimal active  flagella}

\author[Eric Lauga and Christophe Eloy]%
{Eric Lauga$^1$ 
and Christophe Eloy$^{1,2}$}

\affiliation{
$^1$Department of Mechanical and Aerospace Engineering, University of California San Diego, 9500 Gilman Drive, La Jolla CA 92093-0411, USA; \\ 
$^2$Aix--Marseille University, IRPHE UMR 7342, CNRS, Marseille, France.\\[\affilskip]
}

\date{\today}
\begin{document}
\maketitle 
\begin{abstract}
Many eukaryotic cells use the active waving motion of flexible flagella to self-propel in viscous fluids. However, the criteria governing the selection of particular flagellar waveforms among all possible shapes  has proved elusive so far.  To address this question, we derive computationally the optimal shape of an internally-forced periodic planar flagellum deforming as a travelling wave. The optimum is here defined as the shape leading to a given swimming speed with minimum energetic cost. To calculate the energetic cost though, we consider the irreversible internal power expanded by the molecular motors forcing the flagellum, only a portion of which ending up dissipated in the fluid. This optimisation approach allows us to derive a family of shapes depending on a single dimensionless number quantifying the relative importance of elastic to viscous effects: the Sperm number.  The computed optimal shapes are found to agree  with the waveforms observed on spermatozoon of marine organisms, thus suggesting that these eukaryotic flagella might have evolved to be mechanically optimal. 
 \end{abstract}

\section{Introduction}

Many microorganisms swimming in viscous fluids actuate slender appendages, be they flagella or cilia, in a wavelike fashion in order to propel themselves \citep{brennen1977,L1975,childress1981,laugapowers}. The origin of this waving motion rests in the mechanical properties of the surrounding fluid at low Reynolds number. In the absence of inertia, the Stokes equations are  time-reversible and thus any time-reversible deformation of a swimmer or its appendages would result in no average locomotion  \citep{purcell}. To bypass this constraint, microorganisms swim, in general, by using the simplest deformation kinematics indicating a clear direction of time: the travelling wave \citep{lauga2011}.  

As we are all too aware from our efforts at the pool, being able to swim does not however  mean one does it efficiently. In the context of cell motility, a question that has  received some attention in the literature is  the issue of optimal low-Reynolds number locomotion. With infinite degrees of freedom in shape, design, and actuation mechanism, what is the most effective way to self-propel in the absence of inertia? Since the locomotion kinematics in the Stokesian regime scales linearly with the typical actuation frequency of the body, a swimming efficiency needs first to be defined  to  normalise the swimming speed and make the optimality criterion frequency-independent \citep{L1975}. This is typically done by comparing the work done against the fluid to swim (total power) to the work that would be expended by a force dragging the same swimmer at the same speed (useful power). Optimal swimming is then equivalent to either swimming at a fixed speed with minimum  power or swimming at fixed power with maximum speed. 

For some swimmers amenable to precise mathematical or numerical analysis, calculations of optimal kinematics have been proposed. For instance, the optimal swimming gait of a Purcell's three-link swimmer  \citep{purcell} has been characterised by \cite{tam} and \cite{avron08}, as well as  the optimal kinematics of the related three-sphere swimmer  \citep{alouges08}. \cite {shapere87,shapere89_2,shapere89_1} showed analytically that the optimal swimming by surface deformation of spheres and cylinders is achieved by surface waves akin to metachronal waves observed in ciliary fluid transport \citep{brennen1977}, for which the optimal  kinematics has also been studied \citep[e.g.][]{Osterman2011}. 
For swimmers able to impose a tangential velocity on the fluid at their surface without deforming their shapes, a swimming motion referred to as squirming, infinite efficiency can be obtained for infinitely slender swimmers  as shown by \cite{leshansky07}. For a spherical squirmer however, the optimal large-amplitude swimming \citep{michelin2010} or  feeding \citep{michelin2011,michelin2013} problem has a finite efficiency and displays, again,  surface waves.  

Unfortunately, most microorganisms do not have shapes allowing detailed mathematical derivations. The most common shape is that of a cell body, somewhat spheroidal in shape, attached to one or several flagella \citep{braybook}. Eukaryotic cells have active internally-forced  flagella forming planar waves, while bacteria rotate passive helical flagella. The presence of slender flagella allows, physically,  microorganisms to take full advantage of drag anisotropy at low Reynolds number in order to generate drag-based thrust \citep{laugapowers}. In the context of this flagellar swimming, once the issue of optimal body-to-flagella size has been addressed \citep{higdon79a,higdon79b,fujita01}, the important question  becomes how to actuate a flagellum optimally? 

For bacteria locomotion, the optimal shape of  rotated rigid helical  flagella  was  recently derived and found to agree with experimental observations \citep{spa_prl}.  For eukaryotic cells however, the optimal flagellar shape has to be different from the bacterial one since flagella are deformed actively.  A pioneering study demonstrated that, for a given flagellar shape, the optimal instantaneous shape deformation is that of a travelling wave  \citep{piro}. Using a local analysis for the fluid dynamics, Lighthill then derived the optimal shape for a flagellar travelling wave. For an infinite swimmer, he obtained a sawtooth wave \citep{L1975}, an  optimal which remains valid for finite-size swimmers  \citep{piro,pironneau_inbook}.

The optimal flagellar shape thus appears  to be mathematically singular. Yet experimental observations do not show any singularity \citep{brennen1977}. To resolve the discrepancy, different possibilities can be considered. A first possibility could be that   the shapes of eukaryotic flagella have not evolved to be optimal for locomotion, either because locomotion contributes to relatively small energetic costs, or  because other contributions are more relevant biologically. Another possibility could be that cells cannot reach the mathematical optimum because of physical, biochemical, or mechanical constraints. 
A last possibility could resolve the apparent discrepancy: flagella might well be optimal but the energetic costs measured through the power lost in the fluid might not be the pertinent one. The internal structure of an eukaryotic flagellum, called the axoneme, is made up of a small number of polymeric filaments (microtubule doublets) which are caused to slide past each other by the action of a motor protein called the dynein \citep{mbc07}. It is  the action of these molecular motors on the filaments that performs useful work through the consumption of ATP. Only a fraction of this work ends up dissipated in the fluid however, the other fraction being spent on irreversible  bending of the flagellum. Because it quantifies the real biological cost of actuating the flagellum, the work done by these molecular motor is  the correct energetic measure and we believe that efficiency should be defined on this measure to perform adequate optimisation calculations  \citep{eloylauga12}.

In this paper we use this internal energetic measure to derive the shape of the optimal active flagellum. After introducing the physical model for an active, internally-forced periodic planar flagellum in \S\ref{sec:model}, we compute the travelling-wave shape  that maximises the swimming speed for a fixed energetic cost in \S\ref{sec:results}. The shape is a function of a dimensionless Sperm number, $\Sp$, quantifying the relative importance of bending to viscous forces. For finite values of $\Sp$, the  optimal flagellar waveform is  smooth and  becomes singular only  in the hydrodynamic limit, $\Sp\to\infty$. In the elastic limit $\Sp\to 0$, the optimal waveform is composed of  circular arcs with constant curvature of alternating signs. Our optimal shapes are found to agree  with experimentally-measured waveforms of marine microorganisms as discussed in \S\ref{sec:discussion}.

\section{Mathematical model of active flagellum}\label{sec:model}
\subsection{Kinematics}
\begin{figure}
   \centering
   \includegraphics[scale=0.45]{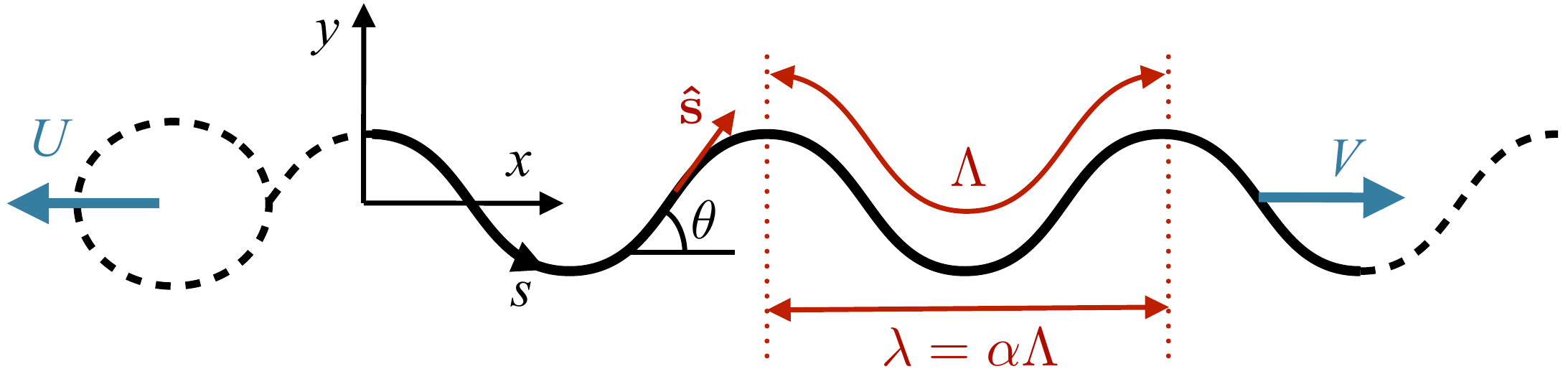} 
   \caption{Mathematical model and notation. We consider an infinite planar flagellum of wavelength $\lambda$ along $x$, with curvilinear coordinate $s$ and tangent unit vectors $\bs$. The wavespeed is denoted $V$ in the $+x$ direction and the flagellum is assumed to be swimming with speed $-U$. The  wavelength measured along the curvilinear direction is $\Lambda\equiv \lambda/\alpha$ ($\alpha \leq 1$).}
   \label{fig:notations}
\end{figure}

We consider an infinite, active, and flexible planar flagellum deforming periodically such that it swims at constant velocity $-U$ in the $x$-direction (figure~\ref{fig:notations}). The flagellum deformation is assumed to be a travelling wave of velocity $V$ and wavelength $\lambda$ along $x$ (or, equivalently, $\Lambda$ along the curvilinear coordinate $s$, with $\lambda=\alpha\Lambda$, $\alpha \leq 1$). The local tangential unit vector is defined as $\bs$.

In the frame moving with velocity $V-U$ compared to the laboratory frame, the deformation of the flagellum appears steady \citep{L1975}. In this frame, the material points on the flagellum move thus necessarily  tangentially  with velocity $-c\,\bs$, where the speed $c$ is such that a material point travels one wavelength $\Lambda$ over a period $T$. Consequently, we have $T=\Lambda/c=\lambda/V$ and thus $V=\alpha c$, where
\begin{equation}
\alpha = \frac{\lambda}{\Lambda}=\frac{1}{\Lambda}\int_0^\Lambda \cos \theta \,\rmd s,
\end{equation}
and $\theta$ is the local angle between the flagellum and the swimming direction (figure~\ref{fig:notations}). The relative velocity between the fluid and the flagellum is therefore given, at any point along the flagellum, by
\begin{equation}\label{eq:bu}
\bu = (V-U)\bx - c\,\bs,
\end{equation}
with a spatial dependence coming implicitly through the variation of $\bs$.

\subsection{Swimming}

In order to compute the fluid forces on the waving flagellum, we use the classical framework of resistive force theory  \citep{Gray1955}, which is the leading-order term of slender-body theory \citep{cox70}. Within this approximation,  the force per unit length exerted by the fluid on the flagellum  can be written as
\begin{equation}\label{eq:bF}
\bF = \xi_\perp \bu + (\xi_\parallel - \xi_\perp) \bs\bs\cdot\bu,
\end{equation}
 where $\xi_\perp$ and $\xi_\parallel$ are the perpendicular and parallel resistance coefficients respectively, and we will further assume that $\xi_\perp=2\xi_\parallel$. Inserting (\ref{eq:bu}) into (\ref{eq:bF}) yields a spatial distribution of force such that
 \begin{equation}
\frac{1}{\xi_\perp} \bF\cdot\bx = (\alpha c-U)(1-\tfrac{1}{2}\cos^2\theta) - \tfrac{1}{2}c\cos\theta.
\end{equation}
To determine $U$, we enforce that the net sum of all hydrodynamic forces  projected along the $x$-direction is zero (free-swimming condition), i.e.~$\int_0^\Lambda\bF\cdot\bx \,\rmd s=0$ \citep{L1975}, leading to  the swimming velocity $U$ given by 
\begin{equation}\label{eq:speed}
U = \frac{1-\beta}{2-\beta}\alpha c, \quad \mbox{with\,\, } \beta = \frac{1}{\Lambda}\int_0^\Lambda \cos^2 \theta \,\rmd s.
\end{equation}
Note that since $c>0$ and $\beta \leq 1$, then $U\geq 0$: a wave travelling to the right leads to swimming to the left.

\subsection{Energetics}

In order to  evaluate the power needed to deform periodically the flagellum, we first need to calculate the active internal torques necessary to the deformation. We use Kirchhoff equations for a flexible rod \citep{Audoly2010} expressing the local balance of forces and moments. In that case, the internal tension, $\bT$, and bending moment, $\bM$, are related to the external force according to
\begin{equation}\label{K}
\bT'=\bF, \quad \bM'+\bs\times \bT+\bq = 0,
\end{equation}
 where primes denote differentiation with respect to the local curvilinear direction $s$, and $\bq$ denotes the  active torque produced by the internal molecular motors. 
Assuming a Hookean constitutive  relation, $\bM = B \theta''\bz$, where $B$ is the  bending rigidity,   the equations \eqref{K} yields an explicit expression for  the internal torque  as
\begin{equation}
\bq = -B \theta''\bz + \bs \times \int_s^L \bF \, \rmd s,
\end{equation}
where $L$ denotes the end of the flagellum or any point with the same phase (in practice any point can be chosen for $L$ since a $s$-shift is equivalent to a time shift). 

The average power needed to perform the deformation is obtained from the scalar product of the internal torque and the local angular velocity
\begin{equation}\label{P}
P = \int_0^\Lambda\left[\bq \cdot \dot\theta\bz\right]^+\rmd s,
\end{equation}
where $\dot\theta = -c \theta'$ is the angular velocity and the notation $[\cdot]^+$ expresses that only positive works are included in the energy budget. In other words, we assume that the flagellum (or more precisely, the molecular motors actuating it) cannot harvest energy from the fluid when the local power given to the fluid is negative. This means that the elastic energy is not conserved and that the work expended  by internal torques is not totally transferred to the fluid, but instead a portion of it is wasted internally due to the irreversibility of internal motors. This assumption is similar to what is classically done when modelling the muscle mechanics, except that, in the present study, the work is done in bending instead of longitudinal compression or extension \citep{Alexander1992}.

\subsection{Swimming efficiency}

The efficiency,  $\eta$, of a given flagellum deformation, $\theta(s)$, is then expressed as the ratio between the power needed to drag one period of the straightened filament in the fluid to the actual power spent to actuate the internal motors
\begin{equation}\label{eq:efficiency}
\eta = \frac{\xi_\parallel \Lambda U^2}{P}\cdot
\end{equation}
This efficiency depends on a single dimensionless parameter, the Sperm number $\Sp$, defined as
\begin{equation}\label{eq:Sp}
\Sp =  \frac{\Lambda}{\ell} ,  \quad \mbox{with }
\ell = \left( \frac{T B}{\xi_\perp}\right)^{{1}/{4}}
\end{equation}
which measures the ratio of the wavelength, $\Lambda$, to an elasto-viscous persistence length, $\ell$. In the traditional approach for optimisation of locomotion at low Reynolds number, the efficiency, denoted $\eta_{\mathrm{fluid}}$ here,  is defined as in \eqref{eq:efficiency} but $P$ is  taken to be the rate of energy dissipated in the fluid, and the elastic nature of the flagellum is not considered \citep{laugapowers}. This was  the assumption made by Lighthill in his optimisation calculation \citep{L1975}. With $P$ defined as in \eqref{P},  the optimisation approach proposed allows us to  rigorously quantify the net  molecular energy expenditure.  As a result, the optimal shape is a function of the elastic modulus of the flagellum through the Sperm number.

\section{Shape of the optimal active flagellum}\label{sec:results}
\subsection{Numerical optimisation}
The optimisation procedure consists in computing, for a given Sperm number $\Sp$, the flagellum shape that maximises the efficiency. This  optimisation problem can be solved numerically by decomposing the flagellum shape onto Fourier modes, such that the local angle is given by
$
\theta(s-ct) = \sum_{n=1}^{N} A_n\cos\left[ 2 \pi n (s-ct) / \Lambda\right],
$
where, in fact, even terms are zero for the optimal shapes due to the problem symmetry.  The swimming velocity can then be determined using (\ref{eq:speed}), and the efficiency using (\ref{eq:efficiency}). From a practical point of view, the Fourier series has been truncated with $N=100$ to have a sufficient spectral resolution, and integrals along $s$ have been discretized onto 1000 elements. The optimisation itself is carried out with \textsc{Matlab} using the sequential programming (SQP) algorithm, starting from a guess value picked as the best efficiency among 1000 random trials. The optimisation calculation is typically ran 20 times for each Sperm number to ensure that the algorithm has reached the global optimum. This approach has been validated by comparing the results for large values of $\Sp$ with the analytical study of \cite{L1975}. He found that, in the hydrodynamic limit where all the work is dissipated in the fluid (equivalent to  vanishing bending rigidity, or infinite $\Sp$),  the maximum efficiency $\eta_{\mathrm{fluid}} = (1-\sqrt{1/2})^2\approx 0.0858$ is reached for a sawtooth shape such that  $\theta = \pm \arcsin(1/(1+\sqrt{2}))^{1/2}\approx 40.06^\circ$.

\begin{figure}
   \centering
   \includegraphics[scale=0.75]{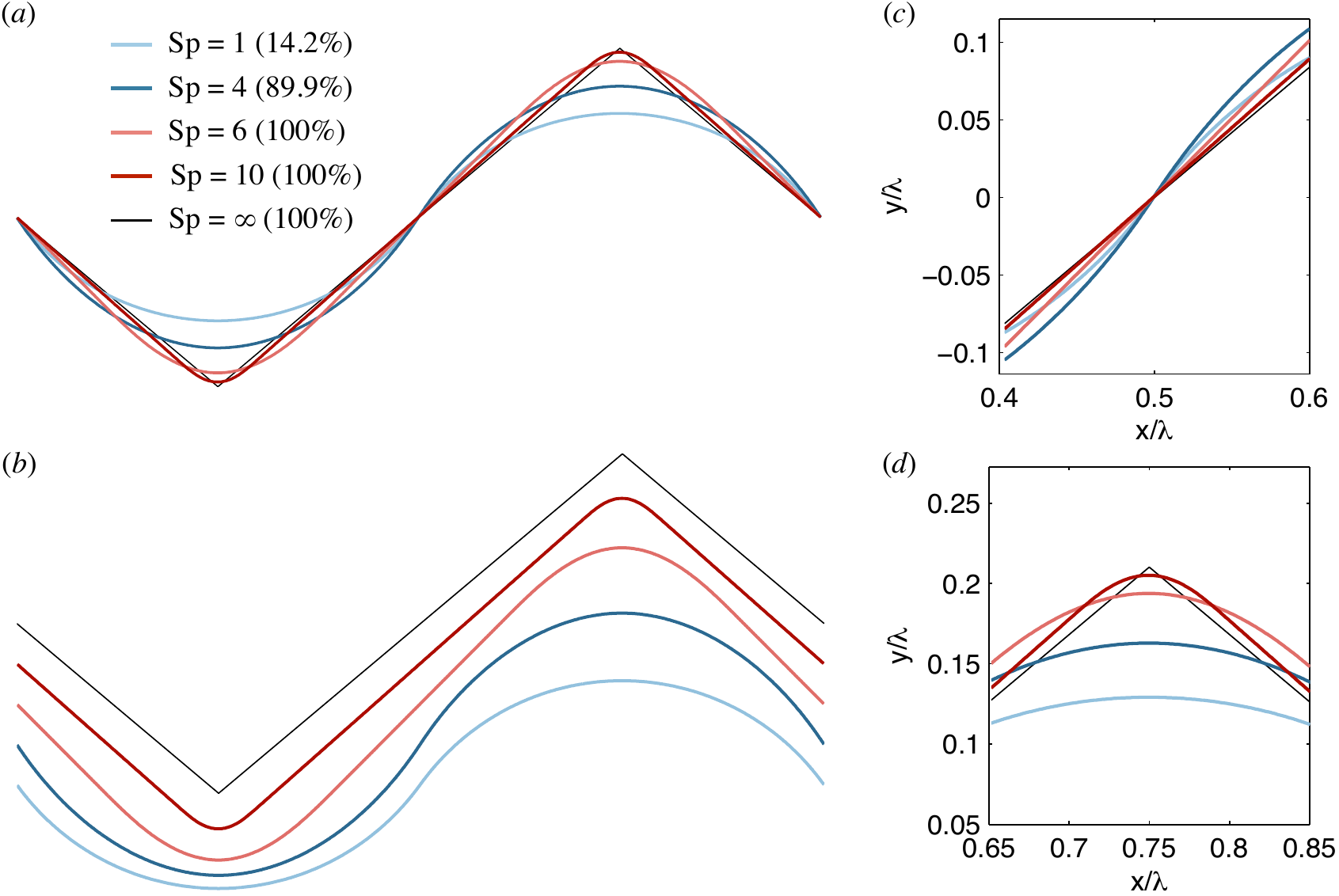} 
   \caption{Energetically-optimal flagellum shapes   for different Sperm numbers, $\Sp$. The sawtooth function, with an angle of $\pm40.06^\circ,$ is the solution of \cite{L1975} and corresponds to an infinite Sperm number (thin black line). The percentages in ({\it a}) refer to the proportion of the power expended by the internal motors (torques) which ends up dissipated in the viscous fluid (see also inset of figure \ref{fig:efficiency}). In ({\it a}) and ({\it b}) we display the overall shapes over a single wavelengths, both superimposed ({\it a}) and shifted ({\it b}). In ({\it c}) we zoom in on the part of the flagellum which intersects  its swimming axis and in ({\it d}) we show a zoom of the region with largest curvature.}
   \label{fig:shapes}
\end{figure}

\subsection{Optimal shapes}
The optimal shapes found numerically by maximising the efficiency at constant $\Sp$ are shown in figure~\ref{fig:shapes}.  As the value of $\Sp$ is increased, the calculated shapes do converge toward Lighthill's sawtooth function.  For finite values of $\Sp$ however, the singular points in $x=\lambda/4$ and $3\lambda/4$ are smoothed out. 

\begin{figure}
   \centering
   \includegraphics[scale=.75]{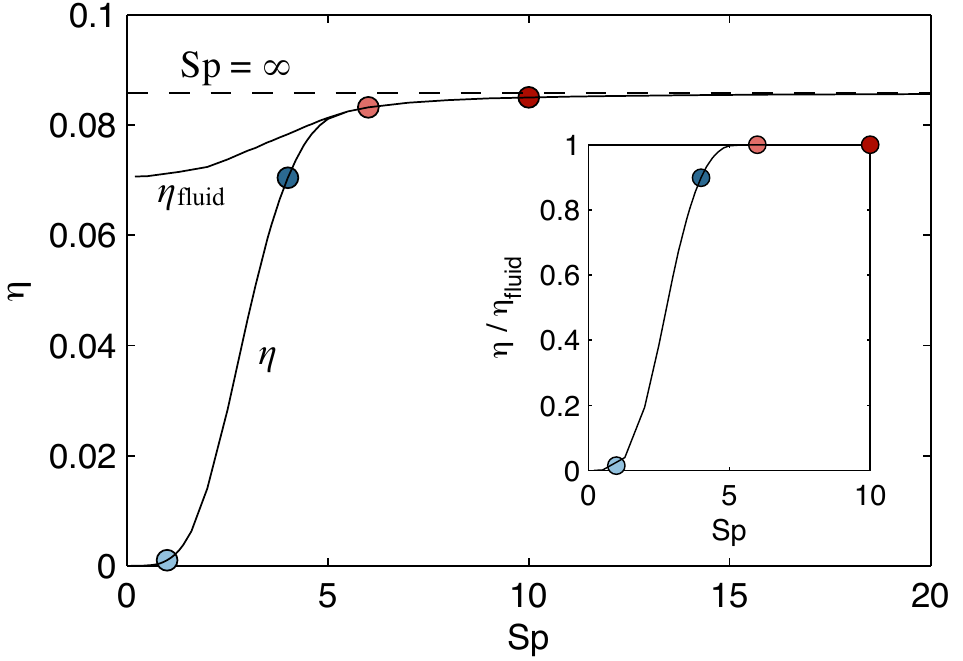} 
   \caption{Efficiency, $\eta$, of the optimal flagellum shape as a function of the Sperm number (line with circles). The coloured symbols correspond to the  cases from figure~\ref{fig:shapes}. The additional solid line shows the efficiency $\eta_{\mathrm{fluid}}$ calculated by considering only the fluid power  while the dashed line shows  the solution of \cite{L1975}, $\eta_{\mathrm{fluid}} = (1-\sqrt{1/2})^2\approx 0.0858$, corresponding to an infinite Sperm number. Inset: proportion of the total internal power  dissipated in the surrounding fluid.}
   \label{fig:efficiency}
\end{figure}

The efficiencies of these optimal shapes are shown in figure~\ref{fig:efficiency} as a function of the Sperm number. We see that the  efficiency is a monotonically increasing function of $\Sp$, which reveals, as expected, that an increase of the bending rigidity leads to an increase of the internal energetic cost.  The external energetic cost, or the energy given to the fluid, is quantified in this figure by also plotting the efficiency $\eta_{\mathrm{fluid}}$ using \eqref{eq:efficiency} but with $P$   equal to the rate of dissipation in the fluid alone.  The ratio between $\eta$ and $\eta_{\mathrm{fluid}}$ corresponds to the ratio between the power given to the fluid and the total power spent and is plotted as an inset in the figure (that ratio is also reported for the five optimal shapes in figure~\ref{fig:shapes}). When $\Sp\lesssim 2.7$, more than half of the power is spent  internally whereas  when $\Sp\gtrsim 5$, the power spent internally is almost negligible (although the optimal shapes continue to depend on the value of $\Sp$).

\begin{figure}
   \centering
   \includegraphics[scale=.75]{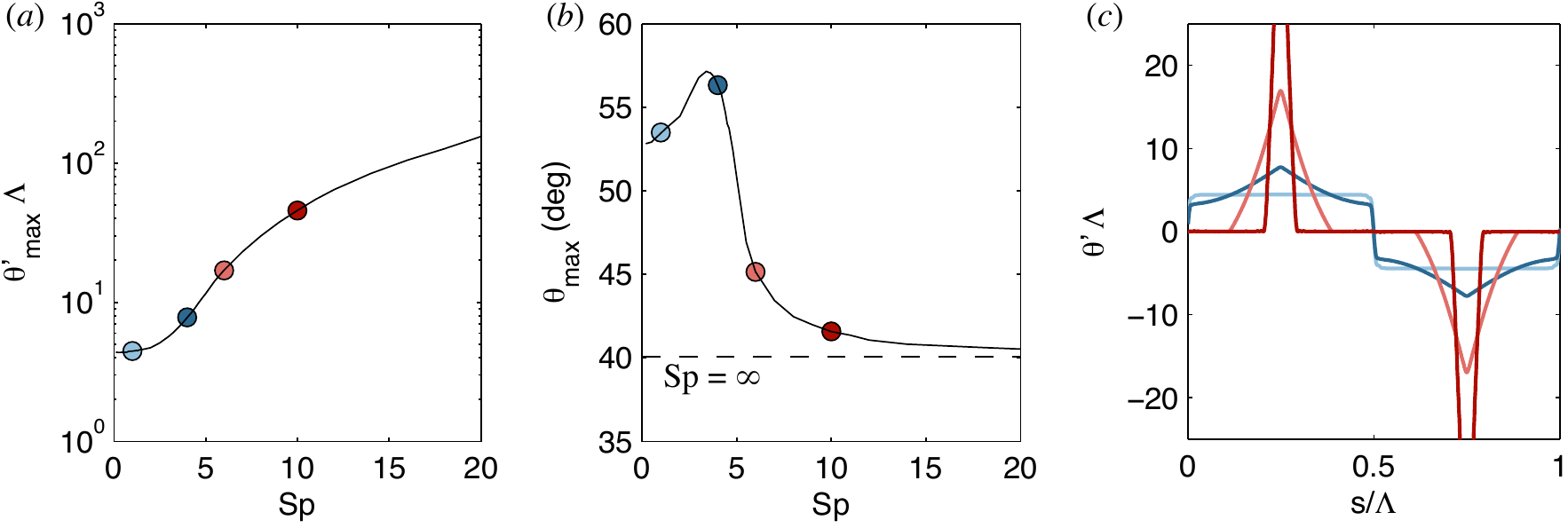} 
   \caption{Curvature and maximum angle of the optimal solutions. (\textit{a}): Maximum curvature (normalised by $\lambda$) as a function of $\Sp$. (\textit{b}) Maximum angle (in degrees),  located at $x=\lambda/2$, as a function of $\Sp$. (\textit{c}) Distribution of dimensionless curvature along the flagellum axis, $x$, for $\Sp= 1$, 4, 6, and 10 (same colour code as in the other figures).}
   \label{fig:analysis}
\end{figure}

In figure~\ref{fig:analysis}, the angle and curvature of the optimal shapes are analysed as $\Sp$ is varied. First, it appears that the maximum curvature of the shape (located in  $x=\lambda/4$ and $3\lambda/4$) is an increasing function of the Sperm number. As anticipated,  increasing the bending rigidity yields a smoother shape. As illustrated in figure~\ref{fig:analysis}\textit{c} however, this monotonic increase of the maximum curvature corresponds to qualitatively different distributions of the curvature along  the flagellum length. 
For small $\Sp$, the curvature is almost constant on the intervals $0<x<\lambda/2$ and $\lambda/2<x<\lambda$ with an abrupt jump (change of sign)  at $x=\lambda/2 $. In that limit, the flagellum shape is an assembly of arcs of circles with identical radii. For $1\lesssim\Sp\lesssim5$, the curvature is no longer 
 constant  on the half-wavelengths, but there is still a jump of curvatures in $x=\lambda/2$. Finally, for $\Sp\gtrsim5$, the jumps disappear and the curvature is zero in two intervals centred around $x=0$ and $\lambda/2$; in this limit of large $\Sp$, the region of  non-zero curvature is concentrated in small intervals  around  $x=\lambda/4$ and $3\lambda/4$, the extent of these region decreasing as $\Sp$ increases.  These different regimes are also apparent  in  figure~\ref{fig:analysis}\textit{b} when looking at the maximum angle $\theta_\mathrm{max}$ between the flagellum centreline and the swimming direction (see also the zoom on the shapes in figure~\ref{fig:shapes}c).  For small values of the Sperm number ($\Sp\lesssim3.4$), this midpoint angle increases with $\Sp$ to reach a local maximum of roughly $57^\circ$, followed by a decrease toward  to the value of $40.06^\circ$ predicted by \cite{L1975} corresponding to an infinite Sperm number.

\section{Discussion}\label{sec:discussion}
In this paper, we have computed the energetically-optimal shape of a flagellum deforming as a pure travelling wave. When the energetic cost is defined as the irreversible power expended by the internal molecular motors actuating the flagellum, this optimisation calculation yields a family of shapes parametrised by a single dimensionless number, the Sperm number, which quantifies the ratio between viscous and elastic effects. When the Sperm number is asymptotically small, the optimal shape is found to be an assembly of circular arcs of constant absolute curvatures. When it is asymptotically large, the optimal shape tends to the sawtooth shape found by \cite{L1975} by considering only the power dissipated in the fluid. 

\begin{figure}
   \centering
   \includegraphics[scale=.75]{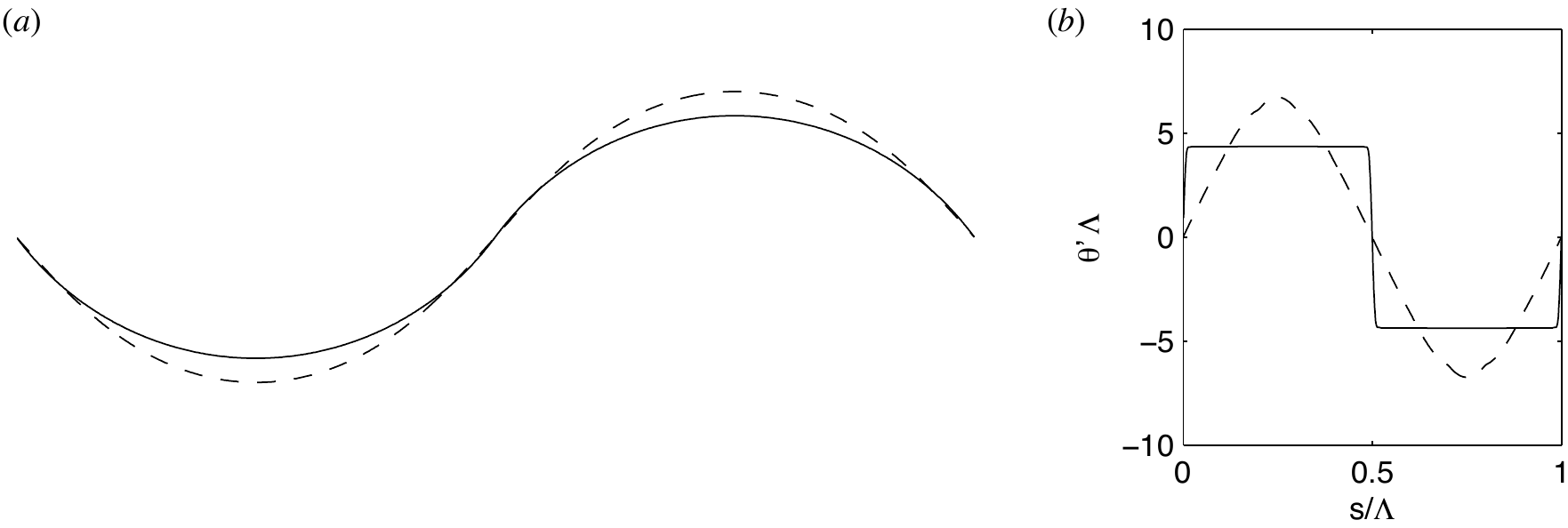} 
   \caption{Comparison between the optimal flagellum shape obtained in the present model (solid line) and the one obtained by \cite{Spagnolie2010} (dashed line) in the elastic regime   (i.e. $\Sp\ll 1$):
   ({\it a})  waveforms;  
   ({\it b}) distribution of  curvatures.
   }
   \label{fig:sav}
\end{figure}

At this point, two comparisons with past work should be made.
First, the calculated shapes can be compared with the results of past optimisation studies. Specifically, we look back at the  study of \cite{Spagnolie2010} who  proposed a physical regularisation of Lighthill's shape by replacing, in \eqref{eq:efficiency}, the rate of viscous dissipation in the fluid  by a linear combination of that dissipation rate and the bending energy stored elastically in the flagellum per period. 
In the limit of strong elastic cost, or equivalently in the asymptotic limit of vanishing Sperm number, our results are compared to the ones of  \cite{Spagnolie2010}  by juxtaposing the optimal waveforms (figure~\ref{fig:sav}a) and the corresponding distribution of curvatures (figure~\ref{fig:sav}b). 
The optimal shape obtained in the present work for vanishing Sperm number is composed of  circular arcs with constant curvature alternating in sign, and has an hydrodynamic efficiency $\eta_{\mathrm{fluid}}\approx 7.1\%$. In contrast, the optimal shape in  \cite{Spagnolie2010}, which is also smooth, is characterised by a sinusoidally-varying shape angle $\theta(s)$ (and thus a sinusoidally-varying curvature) with a smaller hydrodynamic efficiency of $\eta_{\mathrm{fluid}}\approx 6.1\%$. The crucial difference in the analysis between the two studies is that the work of  \cite{Spagnolie2010} penalises all curvature along the flagellum equally through the use of the bending energy measuring the mean square curvature along the shape. As a difference, the present study only penalises location along the flagellum where irreversible work is being done. 

A second comparison can be made with the shapes of actual biological organisms. This can be done both qualitatively and quantitatively. In a famous 50-year-old study,  \cite{brokaw63} described in detail the flagellar waveform of {\it Ceratium}, a marine  dinoflagellate, and reported that contrary to common knowledge   ``the regular form of the wave is not sinusoidal'' but the bent regions are ``circular arcs in which the curvature is constant throughout the bend''. In a followup work analysing the shapes of  spermatozoa of marine invertebrates,  \cite{brokaw65}  similarly found that the observed shapes ``contain regions of constant bending, forming circular arcs, separated by shorter unbent regions''. These results indicate that the optimal shapes found in the present study for small $\Sp$, which exhibit constant curvature, agree qualitatively with those observed experimentally. 

From a quantitative standpoint, we can further compare  the results of the present work with experimental measurements of five flagellar shapes of spermatozoa ({\it Chaetopterus}, {\it Ciona}, {\it Colobocentrotus}, {\it Lytechinus}, {\it Psammechinus}),  and one eukaryotic cell ({\it Tripano\-so\-ma cruzi}), all displaying two-dimensional flagellar beat (Table~\ref{table}), and whose characteristics have been reported in \cite{brennen1977},  \cite{brokaw65}, and  \cite{Gray1955}.  Averaging over all six species, the typical ratio between the flagellar wave amplitude and wavelength is found to be $h/\lambda\approx 0.165$, while the typical maximum curvature is $\theta'_\mathrm{max}\approx 5.75 / \Lambda$. Lighthill's optimal shape gives  $h/\lambda\approx 0.209$ and $\theta'_\mathrm{max}=\infty$. By contrast, in our optimal calculation for $\Sp=4$,  the optimal shapes are characterised by  $h/\lambda \approx 0.163$ and  $\theta'_\mathrm{max}=7.75 / \Lambda$, while for $\Sp=1$,  we obtain shapes with  $h/\lambda \approx 0.128$ and  $\theta'_\mathrm{max}=4.45 / \Lambda$.
What is the value of $\Sp$ that should be used for comparison? 
For spermatozoa, the bending rigidity of the axoneme  is believed to be of the order of $B\approx 2.5\times 10^{-23}$ -- $4.4\times 10^{-22}$\,N\,m$^2$ \citep{Hines1983,gittes93,camalet99}, the typical  period $T\approx 0.04$\,s \citep{brokaw65}, and the drag coefficient  $\xi_\perp \approx 0.003$\,Pa\,s in water \citep[assuming an aspect ratio $\Lambda/a=200$, with $a$ the radius of the flagellum,][]{L1975}. Using equation~\eqref{eq:Sp}, the value of the persistence length is therefore $\ell  \approx 4$ -- $9$\,$\mu$m. Since $\Lambda \approx 28$\,$\mu$m on average for the spermatozoa reported in Table~\ref{table}, the Sperm number of a typical spermatozoon  is $\Sp= \Lambda/\ell \approx 3$ -- $7$.  
Our  optimization approach is therefore able to generate shapes which are close to the experimental observations, suggesting that perhaps eukaryotic flagella are indeed mechanically optimal.

\begin{table}
\begin{center}
\def~{\hphantom{0}}
  \begin{tabular}{@{}rlcccc@{}}
&    						& $\lambda$ 		&  $h$  	& $h/\lambda$&  $\theta'_\mathrm{max}\Lambda$ \\[5pt]
\textsc{Experiments}
& annelid (\emph{Chaetopterus}) 		&  19.5\,$\mu$m		&  3.8\,$\mu$m 	& 0.194 	&  5.93	\\
& tunicate (\emph{Ciona}) 			&  ~\;22\,$\mu$m		&  4.3\,$\mu$m	& 0.195	&  5.45	\\
& sea urchin (\emph{Colobocentrotus})  	&  ~\;30\,$\mu$m 		&  2.8\,$\mu$m  	& 0.093	&  -- 		\\
& sea urchin (\emph{Lytechinus}) 		&  22.6\,$\mu$m 		&  4.6\,$\mu$m	& 0.203	&  5.86 	\\ 
& sea urchin (\emph{Psammechinus}) 		&  ~\;24\,$\mu$m		&  4.0\,$\mu$m	& 0.166	&  --  \\
& protist (\emph{Trypanosoma cruzi})		&  ~3.5\,$\mu$m		&  0.5\,$\mu$m	& 0.142	&  -- 		\\[4pt]
\textsc{Models}
& present model ($\Sp=1$) 			&  $0.856\,\Lambda$	& $0.110\,\Lambda$ &  0.128	& 4.45\\
& present model ($\Sp=4$)			&  $0.806\,\Lambda$	& $0.131\,\Lambda$ &  0.163	& 7.75 \\
& present model ($\Sp=6$) 			&  $0.774\,\Lambda$	& $0.150\,\Lambda$ &  0.194	& 16.9 \\
& present model ($\Sp=10$)  			&  $0.768\,\Lambda$	& $0.158\,\Lambda$ &  0.205	& 45.7  \\
& Lighthill's model ($\Sp=\infty$)	  	&  $0.766\,\Lambda$	& $0.161\,\Lambda$ & 0.209	& $\infty$ \\
\end{tabular}
\end{center}
\caption{\label{table}
Comparison of the wavelength, $\lambda$, amplitude, $h$, amplitude-to-wavelength ratio, 
$h/\lambda$, and the maximum dimensionless curvature, $\theta'_\mathrm{max}\Lambda $ (when available), of  five spermatozoa flagellar shapes and one protist displaying two-dimensional beating \citep[data from][]{brennen1977,brokaw65,Gray1955}, with the results of the present model and Lighthill's optimal shape. }
\end{table}

\begin{acknowledgments}
We thank Mario Sandoval for his help gathering the data in Table~\ref{table}.  We  acknowledge supports from the European Union (fellowship PIOF-GA-2009-252542 to C.E.) and the US National Science Foundation (grant CBET-0746285 to E.L.).
\end{acknowledgments}
\bibliographystyle{jfm}
\bibliography{bib}

\end{document}